\documentclass[useAMS,usenatbib,onecolumn]{mn2e}

\usepackage{graphicx}
\usepackage{amssymb}
\usepackage{natbib}
\usepackage{cite}
\usepackage{textcomp}

\bibpunct{(}{)}{;}{a}{}{,}
\usepackage{hyperref}

\begin{document}

\title[The fingerprints of Photoionization and Shock-Ionization.]{The fingerprints of Photoionization and Shock-Ionization in two CSS sources.\thanks{Based on observations obtained at the Gemini Observatory, which is operated by the Association of Universities for
    Research in Astronomy, Inc., under a cooperative agreement with the
    NSF on behalf of the Gemini partnership: the National Science
    Foundation (United States), the Science and Technology Facilities
    Council (United Kingdom), the National Research Council (Canada),
    CONICYT (Chile), the Australian Research Council (Australia),
    Minist\'{e}rio da Ci\^{e}ncia, Tecnologia e Inova\c{c}\~{a}o (Brazil)
    and Ministerio de Ciencia, Tecnolog\'{i}a e Innovaci\'{o}n Productiva
    (Argentina)}
} 

\author[V. Reynaldi and C. Feinstein]{Victoria Reynaldi$^{1}$
 and Carlos Feinstein$^{1,2}$
\thanks{E-mail: vreynaldi@fcaglp.unlp.edu.ar (VR); cfeinstein@fcaglp.unlp.edu.ar (CF)}\\
$^{1}$Instituto de Astrof\'{i}sica de La Plata, CONICET; Paseo del Bosque s/n, La Plata 1900, Argentina\\
$^{2}$Facultad de Ciencias Astron\'{o}micas y Geof\'{i}sicas, UNLP, Argentina}

\date{}

\pagerange{\pageref{firstpage}--\pageref{lastpage}} \pubyear{}

\maketitle

\label{firstpage}

\begin{abstract}
We investigate the ionization state of the Extended Emission-Line Regions (EELRs) around two compact steep-spectrum (CSS) radio galaxies, 3C~268.3 and 3C~303.1, in order to identify the contribution of photoionization and shock-ionization. We perform a new spectroscopical (long-slit) analysis with GMOS/Gemini with the slit oriented in the radio-jet direction, where outflows are known to exist.
The [Ne V]$\lambda 3426$ emission is the most interesting feature of the spectra and the one key to breaking the degeneracy between the models: since this emission-line is more extended than He{\sc ii}, it challenges the ionization structure proposed by any photoionization model, also its intensity relative to H$\beta$ does not behave as expected with respect to the ionization parameter {\it U} in the same scenario. On the contrary, when it is compared to the intensity of [OII]$\lambda3727$/H$\beta$ and all these results are joined, the whole scenario is plausible to be explained as emission coming from the hot, compressed, shocked gas in shock-ionization models. 
Although the model fitting is strongly sensitive to the chosen line-ratios, it argues for the presence of external and strong ionizing fields, such as the precursor field created by the shock or/and the AGN radiation field.

In this paper, we show how AGN photoionization and shock-ionization triggered by jet-cloud interaction work together in these EELRs in order to explain the observed trends and line-ratio behaviours in a kinematically acceptable way.
\end{abstract}



\section{Introduction}

The interaction between the radio jets and the surrounding interstellar/intergalactic medium (ISM/IGM) in radiogalaxies is always a fascinating issue to explore. Revealing if such interactions might trigger the emission in the large-scale nebulae called Extended Emission-Line Regions (EELR) is not an exception. Likewise, the detailed mechanism by which the ionizing radiation from the active nucleus would reach that regions, being previously filetered by the innermost nebular regions such as the Broad-Line Region or the Narrow-Line Region, and excite the gas in the way we observe it, is still far from being fully described. 

Due to their structure, morphology and radio power, the double-lobed, edge-brightened Fanaroff-Riley type II sources \citep[FRII;][]{fr74} are the most suitable candidates to analyze what kind of role the jets play in the excitation and/or ionization of the EELR against the always present AGN ionizing radiation field. FRII radiogalaxies are usually very extended radio sources, where the radio jets can reach tens or even hundreds of kiloparsecs from the nucleus. The scales of the EELRs are usually shorter than the radio lobe extension, despite of being the most extended gaseous systems. Jet-cloud interaction has been searched in several FRII sources, and it has been shown that the mechanism can take place in different scenarios: where the EELR overlaps the lobes, which is the most intuitive scenario \citep{fei99,ros10a,reynaldi13b}, or even where the jet have passed far beyond the EELR structure, in which case the EELR shows spectroscopic and kinematical signs of such interaction although there is no spatial coincidence with the radio plasma \citep*{sol01,reynaldi13a}. Many of jet-cloud interaction fingerprints are hidden in the kinematics of the gas rather than in radio-optical correlation; 2-D velocity maps such as that of \citet{fu06,fu08} are particulary useful to this kind of analysis.

Compact Steep-Spectrum (CSS) sources are young FRII radiogalaxies \citep{deVries97,odea98}; the optical EELR and the radio structure are overlapped likely because of the sources's youth. This characteristic makes them excellent laboratories to find the traces of interaction between both structures: since it is known that the jet is passing through the surrounding gas, the nebula might have fresh fingerprints of such a process. 

The main motivation of this investigation is to find key spectral features that let us identify the level of competition between shock-ionization triggered by jet-cloud interaction and AGN photoionization. We will analyze long-slit spectra of two CSS sources, 3C~268.3 and 3C~303.1, whose integral field spectra \citep*{shih13} suggest that the interaction might be at work. We will go a step further by looking for signatures of both processes (shock-ionization and photoionization) inside the line or line-ratio behaviours; diagnostic diagrams will be analyzed as well. Our study is focused on finding spectroscopical fingerprints of these ionization processes in order to convert them into powerful tools in the study of any other source. The data obtained for this investigation is presented in Section~\ref{section:Data} and their main characteristics are highlighted in Section~\ref{section:Results}. Then, we analyze the spectra and the ionization models in Section~\ref{discussion}. Finally, Section~\ref{conclusions} organizes all the results to explain the way in which the EELRs of 3C~268.3 and 3C~303.1 were excited.

%
%

\section{The Observations: GMOS/Gemini data.}
\label{section:Data}

The data were obtained with the GMOS facility of the Gemini North Telescope as part of two separate programs during 2011 and 2012. The source 3C~268.3 ($z=0.3717$) was observed on April 2011 (GN-2011A-Q-66 program, PI:~C.~Feinstein), and 3C~303.1 ($z=0.2704$) was observed on February 2012 (GN-2012A-Q-18 program, PI:~V.~Reynaldi). The instrument was set up in long-slit mode, the position angle (PA) of the slit was determined by the EELR's orientation: PA=146\textdegree~ for 3C~268.3 and PA=145\textdegree~ for 3C~303.1. The slits's width were chosen depending on the EELR structure, we have used the 0.5'' width-slit for 3C~268.3 and the 1'' width-slit for 3C~303.1. In both cases, the B600-G5307 grating was used; it yields a resolution of 0.9~\AA~px$^{-1}$. The observation of 3C~268.3 was taken in only one exposure of 43~min, but the observation of 3C~303.1 was split in two exposures of 21~min each one. The detector was binned in 2$\times$2 mode, with which the angular resolution is 0.1454 arcsec~px~$^{-1}$. The linear scales\footnote{$H_0$=73~km~s$^{-1}$~Mpc$^{-1}$,$\Omega_\mathrm{mat}$=0.27; $\Omega_{\Lambda}$=0.73} are 1080~pc~px$^{-1}$ for 3C~268.3, and 784~pc~px$^{-1}$ for 3C~303.1, without projection correction.

Data were reduced following the usual steps: bias subtraction, flat-field correction, wavelength calibration and sky subtraction by using the {\sc gemini-gmos} package reduction tasks (versions 1.11 and 1.12beta) within {\sc iraf} (version 2.14.1). We did not performe cosmic ray rejection because the cosmic ray hits are more easily removed from the line-profile during the measurement procedure than from the entire spectrum during the reduction procedure.

\begin{figure}
\begin{center}
\includegraphics[angle=270,width=0.8\textwidth]{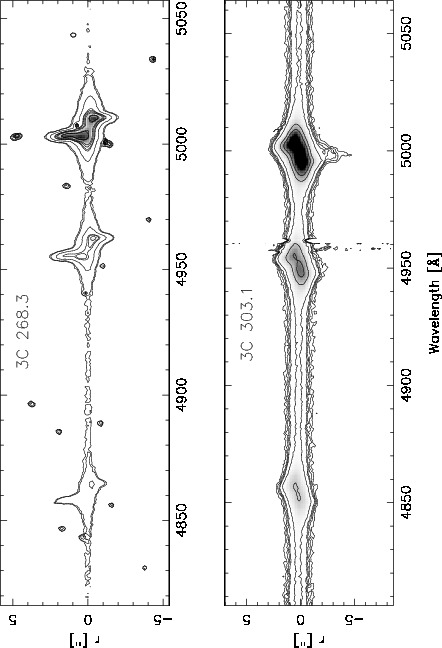}
\caption{Long-slit spectra of 3C~268.3 (upper panel) and 3C~303.1 (lower panel). Only the region between H$\beta$ and the doublet [OIII]$\lambda\lambda4959,5007$ is shown. The contours are plotted to emphasize the line's shapes.}\label{spectra}
\end{center}
\end{figure}

Table~\ref{table-1} lists the set of identified lines in both sources. Our spectral range covers from 3300\AA~ to 5200\AA, approximately. We have arbitrary chosen different distances from the nuclei to show the variations in line-intensities, which are all refered to that of H$\beta$ in each selected position. The line that we have labeled as [NeIII]$\lambda3968$ is actually a blend of [NeIII]$\lambda3967.4$, CaII (3968.44), and HeII(3968.43). The emission coming from this complex was detected in the two of the sources, while in 3C~303.1 we have also found the blend of HI and HeI around 3890\AA. None of these groups take part of our analysis. The two EELR are elongated in the same direction from their nuclei: toward the northwest (NW) and southeast (SE). The distances are measured from the galactic centres, being them identified as the peak of intensity within the continuum emission; hereafter, negative coordinates are used for the NW region and possitive coordinates for the SE region in both objects. The Figure~\ref{spectra} shows the section of the long-slit spectra from H$\beta$ to doublet [OIII]$\lambda\lambda4959,5007$. We have used the grey scale and the contours as well to emphasize the shapes of the emission-lines. The spectrum of 3C~268.3 is plotted in the upper panel and the spectrum of 3C~303.1 is plotted in the lower panel.

%
%

\section{Results.}
\label{section:Results}

Each 1D spectrum was extracted in a pixel-by-pixel way to take advantage of the spatial information across the slit. The [OIII]$\lambda5007$ emission is the most spatially extended feature in both 2D spectra. We have adopted the extension of this emission-line as that of the EELRs. In 3C~268.3 it is extended up to 5.3'' ($\sim$39.3~kpc): 3.3'' to the SE and 2'' to the NW. In 3C~303.1, the EELR is extended up to 8.9'' ($\sim$48~kpc): 3.2'' to the SE, and 5.7'' to the NW. All of the distances are measured with respect to the galactic centres, which were identified as the maximum intensity point within the continuum emission. The continuum emission is very faint in both sources. It is negligible from 0.3'' in 3C~268.3, and from 0.5'' in 3C~303.1. Interstellar extintion correction were not applied since it would require to assume that the gas of the EELRs shares the same physical properties of that of the Milky Way. However, we have calculated the effect that reddening would have over our line-ratios by following the laws of \citet{cardelli89}. These results are shown as reddening arrows in the diagnostic diagrams of Section~\ref{discussion}. The arrows point toward the direction our data would be displaced if the correction were applied.

We derived the velocity fields from the Gaussian decomposition (the Gaussian fitting was fully described in \citealt{reynaldi13a}) of the [OIII]$\lambda5007$ emission-line profile. They are shown in Figure~\ref{VF} together with the maps of full-width at half maximum (FWHM), which let us clearly identify the kinematical complexity inside each nebula. The maps of 3C~268.3 are shown in the left-hand panel, and those of 3C~303.1 in the right-hand panel. Different velocity components were plotted by using different symbols (and colours, available in the on-line version of the journal), which are also used in the FWHM maps. We choosed the PAs of our spectra by following the EELR's optical-UV direction of elongation. These PAs differ from the radio structure's position angles in 15\textdegree~ for 3C~268.3 (PA$_{radio}$=161\textdegree, \citealt{koff96}), and around 14\textdegree~ for 3C~303.1 (PA$_{radio}$=130-133\textdegree, \citealt{koff96,axon00,pri08}). Nonetheless, according to the criteria of \citet{deVries99}, the optical and radio structures of these two CSS sources are globally aligned. Therefore, these velocity fields reveal the kinematics of the gas along (approximately) the radio-jet direction. 

In the following sections, the main characteristics of the whole spectra and the velocity fields are discussed, individually, in detail.

\subsection{3C 268.3}

In 3C~268.3, a clear and extended rotation curve is identified by the (blue) asterisks (Fig.~\ref{VF}, left-hand panels), which stabilizes at around 300-400~km~s$^{-1}$. This is the narrowest component of the fitting, its FWHM is very stable ($\Delta v\sim$150~km~s$^{-1}$) along the entire region. In the innermost, circumnuclear regions ($|r|<1$'') we found the already known disordered motions reported by \citet{holt08}, but there is no sign of high-speed motions even in this circumnuclear region, plotted as (green) triangles. We rather interpret the observed pattern as another, steeper and internal, rotation curve. However, we do detect internal turbulence, which is reflected by the high values of the FWHM: \textlangle $\Delta v$\textrangle$\approx800$~km~s$^{-1}$.

The galactic centre is the region where all the emission-lines are more intense. Line-intensities drop from this zone toward the outskirts in both directions, in a rather abrupt manner. The [OIII]$\lambda5007$ drops to half of its maximum at 0.6'' from the centre, a distance that represents, in contrast, only 17 percent of its total extention. The two highest excitation lines in the observed spectral range are [NeV]$\lambda3426$ (which ionization potential, IP, is almost 100~eV), and HeII (IP=54.4~eV). Both of them are surprisingly extended. HeII is observed in the range -1''$<r<$2'' (or, linearly -7.4$<d<$14.9~kpc), while the [NeV] emission is only detected toward the SE, in the range 0''$<r<$2.5'' (0$<d<$18.6~kpc). This means that, in the region where the two species coexist, [NeV] is more extended than HeII.

\subsection{3C 303.1}

The velocity field of 3C~303.1 is intriguing (Fig.~\ref{VF}, right-hand panels). Although we observe stable motions within each region, it is very difficult to identify a rotation curve. A similar result was obtained by \citet{odea02} by using STIS/HST long-slit spectrum with a slightly different PA (PA=151\textdegree). 
We identify three kinds of kinematical components, namely: the most extended, the internal, and the highest-velocity components. The most extended component is plotted as (red) asterisks; the FWHM ($\Delta v$) ranges between 200 and 700~km~s$^{-1}$. The internal component, plotted as (green) triangles, is also the most turbulent one: $900\lesssim \Delta v \lesssim1200$~km~s$^{-1}$; it is detected mostly toward the SE (-0.5''$<r<$1.5''). Finally, the highest-velocity ($V\sim400$~km~s$^{-1}$) component, plotted as (blue) dots, is found in the region $|r|<2$'' and shows the lowest level of turbulence: $200\lesssim \Delta v \lesssim400$~km~s$^{-1}$.

The central region shows the highest-amplitude velocity among components. The most perturbed motions are observed within the circumnuclear region ($|r|<1$''), where high-velocity movements are detected. None of the observed characteristics in this velocity field are compatible with the expected kinematics within the galactic potential \citep{tad89,odea02}.

The EELR in 3C~303.1 is more extended toward the NW, but the emission-lines are more intense to the SE, which also coincides with the most perturbed region. The highest excitation lines are [NeV]$\lambda3426$ and HeII, as it was the case of 3C~268.3. The former is only detected in the innermost SE region, up to 1.2'' (7.1~kpc); it is also present in the central region, but absent from 0.3'' (1.6~kpc) toward the NW. The emission from HeII, in turn, is slightly more extended and it is observed in the range -1.2''$<r<$1.6'' (or linearly, -6.5$<d<$8.6~kpc). The turbulent region (-2.5''$<r<$2.5''; or $|d|<$13.5~kpc) concentrates the emission of all the species, but it also comprises the entire [OII]$\lambda3727$-zone. Outside this range, only [OIII]$\lambda5007$ is detected.\\

The main characteristics of the two velocity fields are similar to those reported by \citet{holt08} although their spectra were obtained with different PA (the differences with respect to ours are 11\textdegree~ in the case of 3C~268.3 and 15\textdegree~ in the case of 3C~303.1). The outflows that they found in the inner regions ($|r|<1$'') are also reported here. Concerning the outer regions ($|r|>1$''), 3C~268.3 shows no sign of turbulence nor perturbation but clear stability. In fact, the main, dominant and extended component that arises from our Gaussian decomposition plays the role of the narrow component that \citet{holt08} associate to the stationary medium of the galactic halo. In the case of 3C~303.1, the turbulence observed in the inner (circumnuclear) region is still present in the extended gas, up to distances of $r\sim2$'' in both NW and SE directions. 
%
%

\section{The Ionizing Mechanisms: Discussion.}
\label{discussion}

In the last years, several analysis of the ionizing mechanisms taking place in these sources have been made by \citet{labiano05,holt09} and \citet{shih13}. Many of them have proposed that a combination of the AGN ionizing radiation and powerful shocks triggered by the jets might be able to explain the observed ionization state. Different kinds of AGN photoionization models have been already discussed in the former two articles, whose conclusions ruled out the single density models as the main ionizing mechanism. They showed that the spectral characteristics are better reproduced by a model consisting of a mixture of two gaseous systems that differ in their optical-dephts (this same model, among others, will be tested against our observations in this Section). Likewise, the most recent analysis \citep{shih13} suggested that it is likely that some regions have undergone shock processes that have altered the physical condition of the gas, so as to make AGN photoionization more efficient. Our longslit spectra give us the advantage of studying the state of the gas in the jet direction, where shocks are being triggered. In the following, we are going to analyze the main spectral features, the behaviour of the most important species together with models's prediction for both radio sources. One of our most interesting results concerns the behaviour of the highest excitation lines. We will begin our discussion by analysing them under photoionization and shock-ionization predictions; then, we will look for the hints of the ionization processes they could hide. Shock-ionization will be studied by using the Mapping-{\sc iii} library \citep{all08,gro10} with particular attention to the role that the local ionizing field plays. AGN photoionization will be analyzed through the {\it mixed-medium} models of \citet{bin96,bin97} which propose that the medium is composed by a mixture of matter-bounded (MB) and ionization-bounded (IB) clouds; the kinematics of the extended gas will also take part of the study. 

\subsection{The behaviour of [Ne V]$\lambda3426$.}\label{NeV}

The main goal of the {\it mixed-medium} AGN photoionization models is their ability to reproduce the intensity of high excitation/ionization emission lines. Four models were developed; the first of them \citep[][hereafter model T]{bin96} established a new excitation sequence. The $A_{M/I}$ sequence replaces (in these models) the ionization sequence given by the ionization parameter U\footnote{The dimensionless ionization parameter U is defined as $Q/4\pi r^2 n_e c$, being $Q$ the amount of ionizing photons emitted by the AGN per unit time, $r$ the distance between the source and the cloud and $n_e$ the cloud's number density.}, taking into account that the nebula is composed by the optically thin MB clouds and the optically thick IB clouds. The three models of \citet[][they named them H, M and L; this nomenclature is kept in the following]{bin97} made use of the main results of the former model, but these new models were also tuned to reproduce coronal lines such as [FeVII]$\lambda6086$. The four models mainly differ on the ambient density and the (both initial and higher) ionization parameter U of the MB clouds; the whole set of parameters are listed in Table~\ref{table-2}.

In order to be able to simultaneously reproduce these coronal lines and low excitation lines, it is mandatory to consider that different values of the ionization parameter U (hidden in the MB-IB clouds mixture) characterize the ambient medium \citep{stas84,bin96,kom97}. With this picture in mind, the zones with the highest U are responsible for the emission of the highest excitation lines, such as the aforementioned [FeVII]$\lambda6086$ or [NeV]$\lambda3426$ which is originated in similar physical conditions than [FeVII] \citep{stas84}. The fact that the highest excitation emission line in the spectra of both sources 3C~268.3 and 3C~303.1 is, indeed, [NeV]$\lambda3426$ let us analyze the H, M and L models with this information: since they were developed to reproduce the intensity of [FeVII], and [FeVII] originates in the same conditions than [NeV], then we can expect that these models also reproduce [NeV].

In Figure~\ref{neon} (a coloured version of this figure is available in the online version of the journal) the behaviour of neon lines [NeV]$\lambda3426$/[NeIII]$\lambda3869$ is shown against the excitation axes [OIII]$\lambda5007$/H$\beta$ \citep[right-hand panel;][]{bal81}, and HeII/H$\beta$ \citep[left-hand panel;][]{bin96,bin97}. Hereafter 3C~268.3 is plotted as circles (magenta) and 3C~303.1 is plotted as squares (light blue). The bigger circle and square represent the line-ratios in the galactic centres's position, respectively. We have plotted the four AGN photoionization models as well as the self-ionizing (i.e. shock~+~precursor) ionization model (solar abundances; the ambient preshock density is $10^2$~cm$^{-3}$). The transverse magnetic field is combined with the ambient density to form the magnetic parameter $B/n^{1/2}$ \citep{dop95,dop96}. The shock velocities are also shown, in the range 200-1000~km~s$^{-1}$.

According to our prediction, photoionization models provide a global better fit to the observations. Furthermore, not only they reproduce the observed intensity in neon's ratio but their trend as well; note (from the position of the biggest symbols) that the intensity of [NeV]$\lambda3426$/[NeIII]$\lambda3869$ increases from the galactic centre toward the outskirts in each direction for both sources. The fittings are better in the high-values-tail of the $A_{M/I}$ sequence, which represents a MB-dominated medium. It also means that the usual ionization parameter U takes its highest values there \citep{bin96,bin97}. In the right-hand panel we can observe that the best fittings are obtained with the T and H models, the two models for which the MB clouds are the most excited \citep[i.e. the highest U were set to the MB systems in model T and H; ][]{bin96,bin97}.

\begin{figure}
\begin{center}
\includegraphics[angle=270,width=0.7\textwidth]{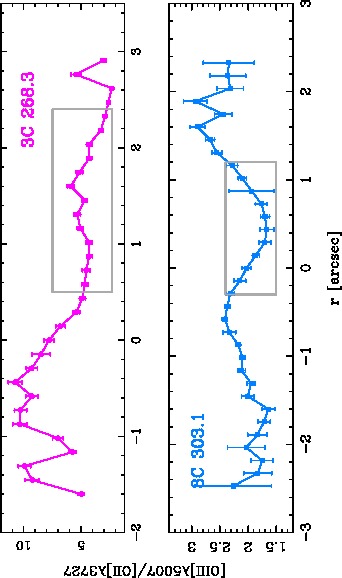}
\caption{The ionization parameter U expressed through its relationship with [OIII]$\lambda5007$/[OII]$\lambda3727$ \citep{pen90} as a function of distance from the nuclei: 3C~268.3 in the upper panel, and 3C~303.1 in the lower panel. The boxes show the regions where [NeV]$\lambda3426$ emission is detected. The emission regions toward the NW are expressed with negative coordinates, while possitive coordinates indicate the SE regions in both galaxies.}\label{U}
\end{center}
\end{figure}

\begin{figure}
\begin{center}
\includegraphics[angle=270,width=0.8\textwidth]{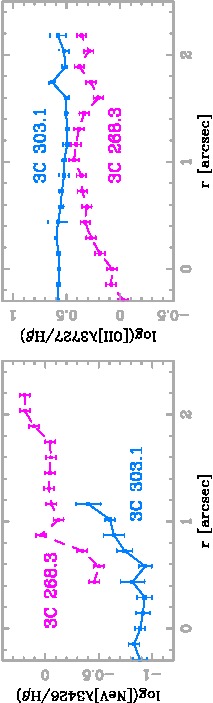}
\caption{Spatial variation of [NeV]$\lambda3426$ (left-hand panel) and [OII]$\lambda3727$ (right-hand panel) relative to H$\beta$ in the regions where we have observed the decreasing U (the boxes in Fig.~\ref{U}).}\label{oxs}
\end{center}
\end{figure}

Since the nebulae seem to be MB-dominated systems, let's assume as a first-order approximation that the EELRs have no mixed-matter (that is, we will assume that the IB component is negligible) to be able to evaluate the behaviour of the ionization parameter U in these two nebular regions as a function of distance from the nuclei. We are interested in knowing what kind of trend U follows in the whole nebulae, and particulary in the region where [NeV]$\lambda3426$ is found. Figure~\ref{U} shows such a behaviour through the relationship between U and the oxygen ratio [OIII]$\lambda5007$/[OII]$\lambda3727$ ($log(U)=-2.74+log(\textrm{[OIII]}\lambda5007/\textrm{[OII]}\lambda3727)$; \citealt{pen90}); 3C~268.3 in the upper panel, and 3C~303.1 in the lower panel. Distances are expressed in arcsec, where $r=0$ represents the nuclei; the NW regions are expressed in negative coordinates, and the SE regions are expressed in possitive coordinates. From these plots we can see that the two ionization parameters U do not show the expected geometrical dilution, the typical pattern of central photoionization, neither they peak at the centre. On the contrary, U increase in off-nuclear regions. In 3C~268.3, U grows up toward the NW and there is also a secondary maximum at 1''-2'' to the SE. In 3C~303.1, U reaches its maximum value in the SE region, at 1''-2'', and also increases toward the NW in the circumnuclear region (0.5''). The (grey) boxes show the regions where [NeV]$\lambda3426$ emission is detected. The spatial behaviours of [NeV]$\lambda3426$ (normalized to H$\beta$) is these same regions are plotted in Figure~\ref{oxs} (left-hand panel; 3C~268.3 as dashed line, and 3C~303.1 as continuous line). It is straightforward to note that the [NeV]$\lambda3426$ emission is very extended: in 3C~268.3 it is observed up to 2.5'' ($\sim$18.5~kpc) to the SE, a distance that represents 75 per cent of the EELR's extension (given by  [OIII]$\lambda5007$). In 3C~303.1, [NeV]$\lambda3426$ emission is observed up to 1.2'' ($\sim$6.5~kpc), i.e. almost a third of the EELR-SE's extension. This is worth noting because [NeV] is the highest-excitation emission-line in the two spectra, which needs energy around $\sim$100~eV to be produced. But the most interesting behaviour is that the [NeV]$\lambda3426$ emission increases with distance. And even more interesting, and contrary to photoionization predictions, it increases over the regions where U decreases. The boxes in U-panels (Fig.~\ref{U}) are shown to emphasize this trend. We have obtained the mean values of U (with the aforementioned relationship between U and [OIII]/[OII]) over the [NeV] emission regions, so we can compare them with models's prescription: $\langle U_{3C268.3} \rangle =7\times 10^{-3}$ and $\langle U_{3C303.1} \rangle =4.2\times10^{-3}$. The ionization parameters that produce bulk of [NeV] emission in the T and H models (this is within MB clouds) are among one and two orders of magnitude higher than these values \citep{bin96,bin97}. So, the characteristics of [NeV] emission present manifold issues: it is surprisingly extended, so very energetic photons must be present at very large distances from the ionizing source; its intensity increases with distance while photoionization predicts it should decrease with distance; the ionization parameter required to form this emission-line should be one or two orders of magnitude higher than observed. The direct consequence of all mentioned behaviours is that [NeV] emission should vanish while U decreases, but against photoionization predictions we observe that [NeV] emission increases its intensity in the regions where U decreases.

\subsection{The fingerprints of Photoionization and Shock ionization.}

\citet{deVries99,labiano05,holt09,shih13} have suggested that a combination of AGN photoionization and shock-ionization, triggered by jet-cloud interaction, might be responsible for the EELR's emission in both sources. \citet{odea02} showed that the kinematics of 3C~303.1 is driven by shocks, and latter \citet{labiano05} concluded that the nucleus of 3C~303.1 do not produce enough photons to power the emission line luminosity in the extended gas. \citet{holt09} have studied 3C~303.1 and 3C~268.3 as well; they have analyzed both single-density and multi-phase photoionization, and  also shock-ionization models throughout several diagnostic diagrams. There, it was shown that, on the basis of the location of both the data and the models in each diagram, mixed-medium photoionization and shock-ionization are capable of producing the observed line-intensities. And recently, \citet{shih13} analyzed these EELRs with the integral field spectroscopy facility (GMOS/IFU) on Gemini North and found that both mechanisms are equally important in the nebulae, although in some places one dominates the ionization state and in other places they work together. 

\begin{figure}
\begin{center}
\includegraphics[angle=270,width=1\textwidth]{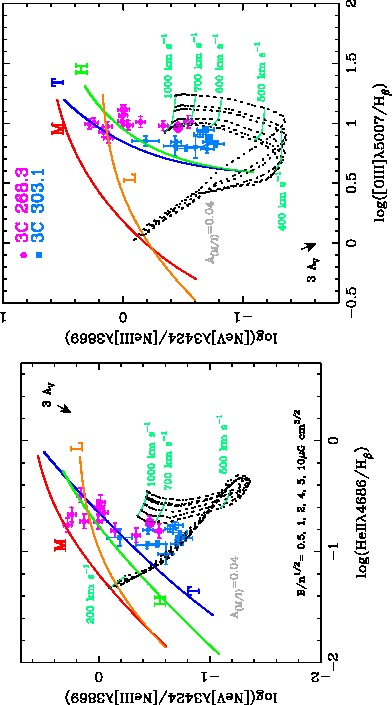}
\caption{Analysis of [Ne{\sc v}]$\lambda3426$, the highest excitation emission-line in the spectra of 3C~268.3 (circles, magenta) and 3C~303.1 (squares, light blue) throughout the [Ne{\sc v}]$\lambda3426$/[Ne{\sc iii}]$\lambda3869$ ratio, under photoionization and shock-ionization predictions. He{\sc ii}/H$\beta$ and [O{\sc iii}]$\lambda5007$/H$\beta$ are used as excitation axes (left-hand panel and right-hand panel, respectively). The four photoionization models (T, H, M, L) are plottes as solid lines; the parameter $A_{M/I}$ increases from left to right. The shock models are plotted as dotted lines (black, one line per magnetic parameter $B/n^{1/2}$); thin solid lines (green) show the shock velocities. Open symbols represent the NW regions and filled symbols represent the SE regions in both galaxies.}\label{neon}
\end{center}
\end{figure}

Here we want to find the traces of shock-ionization and photoionization by using the information hidden inside both the line-intensities and line-ratio. The behaviour of [NeV]$\lambda3426$ is the first step to comprehend to which extent the two mechanisms are related. As we have discussed for this emission-line,  the best fitting in diagnostic diagrams is achieved for the {\it mixed-medium} photoionization models (Fig.~\ref{neon}), but their key parameters fail in reproducing the physical conditions where this line should be formed when observations are confronted with theory. So the question remains open: how it is possible to have such a high-excitation line extended up to so large distances from the radiation source, where U is so low. However, if we look into shock-ionization models we find that the answer might be hidden in the shocked gas.

When the gas is shocked, it reaches very high temperatures and it is compressed giving place to a significant reduction to the ionization parameter U. Such a drop in U, which is produced in the recombination region behind the shock front, enhances the emission of low excitation lines, as it is the case of [OII]$\lambda3727$ \citep{dop95,dop96,all08,stas09}. In order to further explore this scenario, we have plotted the spatial variation of [OII]$\lambda3727$ (normalized to H$\beta$) in the regions where we have observed the decreasing U. The right-hand panel of Fig.~\ref{oxs} shows these trends in both galaxies (again, 3C~268.3 as dashed line, and 3C~303.1 as continuous line). The chosen range in the x-axis is also the region where the [NeV]$\lambda3426$ emission is observed. From this plots we verify that [OII]$\lambda3727$ emission is enhanced where U decrases. Altough it is more relevant in the case of 3C~268.3 than in 3C~303.1, the \textquotedblleft compression-hypothesis'' is valid.

This is an important result, but it is not conclusive at all. Even when the increase in [OII]$\lambda3727$ emission and the drop in U with growing distance might have the same explanation, the issues concerning [NeV]$\lambda3426$ remain unanswered. The [OII]$\lambda3727$ is a low-excitation emission-line, which intensity is enhanced by the compression of the shocked gas (the same effect that lowers U), but [NeV]$\lambda3426$ is not. And a new question opens: since the recombination region will exist no matters whether the shocks are self-ionizing or not, how can we know what kind of shock processes are taking place in these EELRs? 

The answer to this latter question can be easily found in the Mapping-{\sc iii} library \citep{all08} by comparing the predictions of pure shock and shock + precursor (self-ionizing) models, which will be discussed in the next paragraphs. Concerning the [NeV]$\lambda3426$ emission in particular, shock models can also explain why high-excitation emission-lines are so efficiently formed in a region where U decreases. This happens just behind the shock front (closer to the shock front than the [OII]-emitting zone) when the shocked, compressed and therefore very hot gas is also exposed to {\it any} radiation field, no matters whether this ionizing field is the one that comes from the AGN, or from the shock itself, or even both. Most of the soft X-ray and extreme ultraviolet (EUV) fluxes are emitted in this thin layer of gas, where also collisional ionization becomes relevant due to the high temperature \citep{shapiro92,rob02,all08,stas09}. Furthermore, if the medium is magnetized (i.e. if there is a non-negligible magnetic field in the nebula), the effects of that or those radiation fields become stronger because the magnetic field acts to limit the gas compression \citep{shapiro92,dop95,dop96,all08}. And even more, the radiative decay that drives the emission of [NeV]$\lambda3426$ photons is not as affected by gas compression as the other optical lines, given the high transition's critical density ($1.6\times 10^7$~cm$^{-3}$, almost two orders of magnitude higher than [OIII]$\lambda5007$; \citealt{pet97}).

\begin{figure}
\begin{center}
\includegraphics[angle=270,width=1\textwidth]{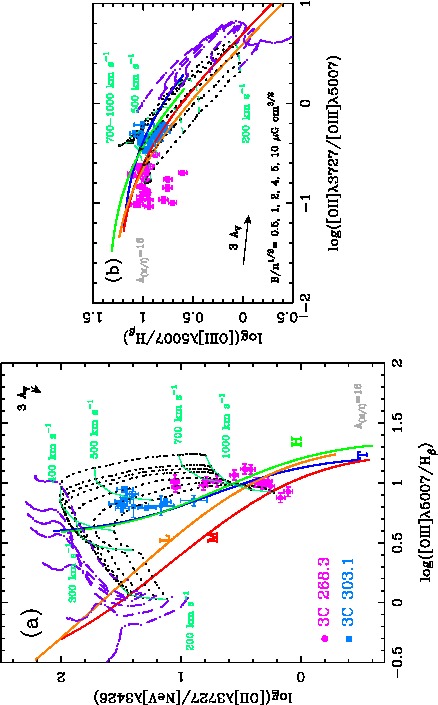}
\vspace{2mm}
\includegraphics[angle=270,width=1\textwidth]{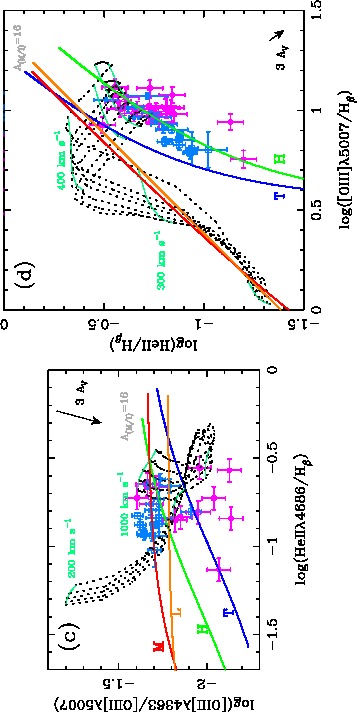}
\caption{Diagnostic diagrams formed with the most important emission-lines in the spectra of 3C~268.3 (circles, magenta), and 3C~303.1 (squares, light blue). As in Fig.~\ref{neon}, open symbols represent the NW regions and filled symbols represent the SE regions in both galaxies. Mixed medium photoionization models and self-ionizing shock~+~precursor models are the same as in Fig.~\ref{neon}. The dot-dashed lines (violet) in diagrams I and II represent the pure shock models.}\label{diagrams}
\end{center}
\end{figure}

We have studied the pure-shock and self-ionizing shock models in order to find out the relevance of the local field created by the shocks onto line-ratio intensities. The Figure~\ref{diagrams} joins a set of diagnostic diagrams where we have plotted the observations (the same references as in previous figures), the four photoionization models (solid lines), and the two groups of shock models (only plotted in the upper two diagrams). The self-ionizing shocks, those shocks that create the HII-like precursor region, are plotted with dotted lines (black); the pure-shock models are plotted as dot-dashed lines (violet). We have combined low-, medium-, and high-excitation lines; the [OIII]$\lambda5007$/H$\beta$ and HeII/H$\beta$ line-ratios have been used again as excitation axes. 

\begin{figure}
\begin{center}
\includegraphics[angle=270,width=0.5\textwidth]{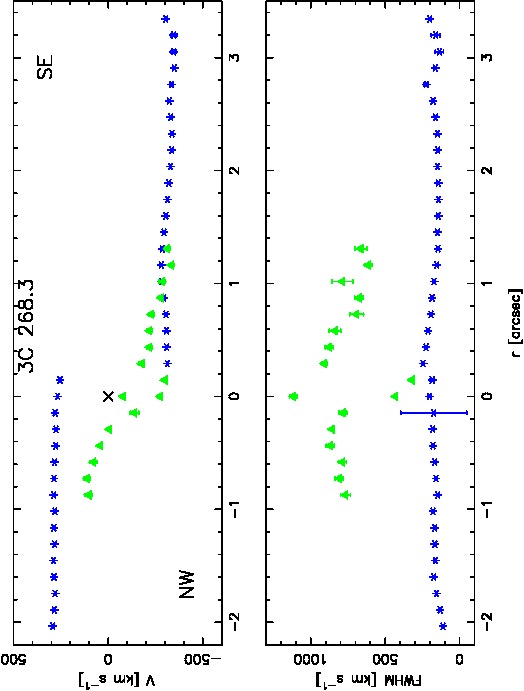}~~~\includegraphics[angle=270,width=0.5\textwidth]{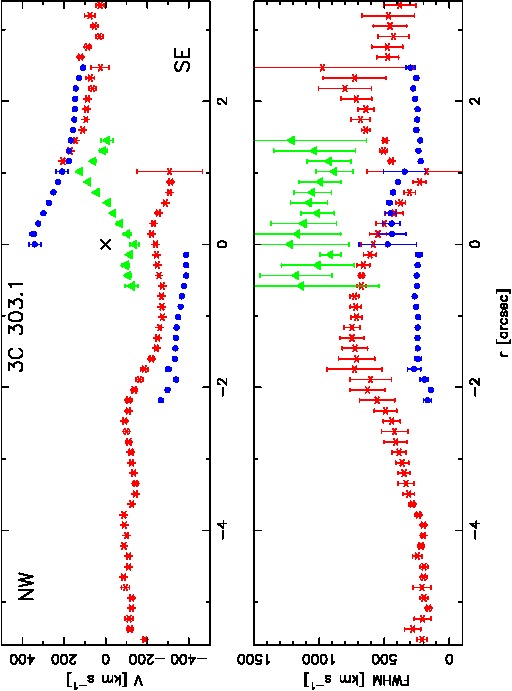}
\caption{Velocity fields (upper panels) with their associated FWHM maps (lower panels) obtained with the Gaussian decomposition of [OIII]$\lambda5007$ emission-line profile in the slit direction. The velocities are referred to the systemic velocity of each galaxy. The galactic centres are shown by crosses at $V=0$. Different symbols are used to separate different kinematical components based on their FWHM. Left-hand panels: 3C~268.3, PA=146\textdegree. Right-hand panels: 3C~303.1, PA=145\textdegree.}\label{VF}
\end{center}
\end{figure}

The diagram in the upper-left corner, (a), involves [NeV]$\lambda3426$ again, but in this case it is combined with [OII]$\lambda3727$. Unlike the [NeV]$\lambda3426$/[NeIII]$\lambda3869$ line-ratio (Fig.~\ref{neon}), the observations of [OII]$\lambda3727$/[NeV]$\lambda3426$ in 3C~303.1 are well explained by both photoionization and moderate-velocity (400-500~km~s$^{-1}$) shock models. The latter fitting is relevant, since the models fit the observations in a velocity range compatible with what we have found in the velocity field (Fig.~\ref{VF}, right-hand panel). The tendency of 3C~268.3 data to be located toward the highest excitation end of both photoionization and shock-ionization is kept. But there is no kinematical evidence to support the shock-ionization fitting as true. The velocity field of 3C~268.3 is remarkably stable (Fig.~\ref{VF}, left-hand panel), showing almost no sign of perturbation. The situation is quite different for the photoionization models because, given the already discussed relationship between [NeV] and [FeVII], and [FeVII] with the models's intrinsic parameters, it is not surprising to obtain a good agreement among these models and the data when [NeV] is involved. However, we do not discard this fitting at all. We just understand these features must be highlighted to continue analyzing the models in detail. 

The second diagram, (b), in the upper-right corner of Fig.~\ref{diagrams}, emphasises the complexity inherent to the determination of the main ionizing mechanisms in these kind of nebulae. This diagram is useful to show again that the observed line-intensities are globally incompatible with pure-shock model's predictions, while they might represent extreme cases of the other two set of models. The four photoionization models are almost undistinguishable, and they also overlap the shock-ionization models. A fast inspection over all other diagrams indicate that at least one high-excitation line should be present (absent in the diagram b) in any diagnostic diagram to observe different behaviours among the models. The data of 3C~303.1 locate in the overlapping region around 400-500~km~s$^{-1}$ shocks. The observations of 3C~268.3 in turn, lie outside any kind of prediction. However, this diagram involves the [OII]$\lambda3727$/[OIII]$\lambda5007$ line-ratio, strongly related to the ionization parameter U \citep{pen90}, whose \textquotedblleft unconventional'' pattern was already discussed.

The contribution from the precursor must be present, which is equivalent to affirm that the shocks are, indeed, creating a local ionizing field capable of changing the state of the surrounding gas. The photons travelling upstream the shock create the HII-like precursor region, but the injection of energy in the recombination region caused by the photons that travel downstream the shock enhances the emission of high-excitation lines in the UV, such is the case of [NeV]$\lambda3426$. However, we see that the precursor's strength is not enough to explain the observed intensity of [NeV]$\lambda3426$, and, at lower level, HeII too. Since the precursor contribution is not negligible we affirm that photoionization processes are specially important. The term 'photoionization' is deliberately used here to encompass the ionization by the AGN central field and the ionization from the shock's local field; and also to highlight how difficult is to establish where the photons actually come from. 

These two diagrams let us comprehend which is the effect of the external ionizing field onto the predicted line-ratios (line-intensities). The presence of the external (but locally created) radiation field displaces the model's predictions toward the region where the observations are actually located. If we take into account that, in addition to this field, we must consider the presence of the central and very powerful ionizing field from the AGN (despite it has been leaked up to the distances involved), we can figure out that this extra contribution will enhance the prediction's displacements in the same direction as the precursor does. This can be easily verified by using the diagrams, given these displacements occur also toward the region where the highest-excitation edge of $A_{M/I}$ sequence is found. But, since the overlapping occurs for the most violent shocks \citep[line-intensities are also dependent on shock velocity; ][]{dop95,dop96,all08}, the velocity fields become particularly important tools for rejecting or supporting the shock scenario.

The last two diagrams (c, d) in the bottom of Fig.~\ref{diagrams}, combine medium- and high-excitation lines. For the sake of the analysis we have not drawn the pure-shock models anymore. In the diagram (c, lower-left corner), we have used the (somewhat problematic) [OIII]$\lambda4363$/[OIII]$\lambda5007$ line-ratio. This oxygen ratio is related to the gas temperature and, because of that, has always been controversial for the diagnostic analysis. However, [OIII]$\lambda4363$ contains valuable information when shock processes are present, since the bulk of [OIII]$\lambda4363$ emission forms in the recombination region \citep{dop95,dop96,bin96,all08}. So, it is worthwhile to analyze what happens with this diagram altough we have in mind that the so-called \textquotedblleft temperature problem'' might play an important (and still unknown) role. Even when both set of models predict the observed range in [OIII]$\lambda4363$/[OIII]$\lambda5007$ line-ratio, the data show no trend at all. Regarding 3C~268.3, the data is widely dispersed and they show no tendency in favour of shock-ionization nor photoionization. The observations of 3C~303.1, in turn, seem more concentrated over the moderate-velocity (300-500~km~s$^{-1}$) shock-ionization's predictions; it was also the case in the first diagram (a), despite this latter fitting is less significant than that.

In the last diagram (d, lower-right corner), we have plotted one excitation axis against the other, and we apparently obtain the best fitting for the photoionization model in the two sources. Concerning shock-ionization, 3C~268.3 is again located over the region of very fast shocks which has no kinematical support to be considered as a valid fitting. 3C~303.1 is completely outside the shock-ionization predictions because the [OIII]$\lambda5007$/H$\beta$ intensity is higher than expected for the observed HeII/H$\beta$ intensity (data are displaced to the right side of the diagram). From the photoionization perspective, we would be obtaining the best fitting, however the caveat that the main parameters of the photoionization models fail in reproducing the observed physical condition in the nebulae should be remembered.

There are some common characteristics to all the six diagrams (Figs.~\ref{neon}~and~\ref{diagrams}) that are worth highlighting at this point: every time that at least one high-excitation line ([NeV]$\lambda3426$, HeII) is used, the observations of 3C~268.3 tend to be fitted by photoionization models (the diagram (c) is an exception), and those of 3C~303.1 tend to be located over the region in-between the models, or over the overlapping region (the diagram (d) is a clear exception). But when the diagrams are formed with medium- and low-excitation lines ([NeIII]$\lambda3869$, [OIII]$\lambda5007$ in the former group, and [OII]$\lambda3727$, H$\beta$ in the latter) self-ionizing, recombination-region-dominated, shock models reproduce the observations of 3C~303.1 in a better way. In contrast, the apparent fitting of shock-ionization to 3C~268.3's observations in the region of very fast shocks should be ruled out because of the lack of kinematical evidence to support such an scenario. Despite this discrepancy, it is very likely that shocks be at work in 3C~268.3 too, at least at a lower level.

\section{The Whole Picture: Summary and Conclusions.}
\label{conclusions}

Several authors before us have proposed that AGN photoionization and shock-ionization might be working together in these two CSS sources. It is known that in the shock-ionization scenario, at the distances we are looking to, photoionization from the AGN might be masked out by ionization processes triggered by the shocks. 

We start this section by saying that none of the models are able to completely explain the ionization state in the two EELRs by themselves. Diagnostic diagrams's results are tricky and strongly dependent upon the chosen line or line-ratios. The main challenge of our study was to explain the (apparently inconsistent) behaviours of low- and high-excitation lines simultaneously. These behaviours hide the fingerprints that both processes have left in the nebulae.

We have analyzed key emission lines along the entire nebulae (in the slit direction) to test them against model's predictions, and we have also studied a wide sample of line-ratios. We have seen how shock-ionization predictions are displaced within each diagram (with respect to pure-shock models) when the contribution from the precursor, created by the shocks, is taken into account. It was also shown that the highest excitation tail of AGN mixed-medium photoionization sequences (predictions) overlap and/or even overpredict the highest shock+precursor models's predictions. So, the diagrams have a highest excitation zone (HEZ), and the HEZ is shared by the two set of models. The way in which these models behave let us conclude that, if both shock-ionization and AGN photoionization act together, then the presence of the strong AGN ionizing field accentuates the displacements of shock-ionization's predictions in the direction of the HEZ.

Altough it is almost impossible to quantify each contribution individually, we conclude that 3C~268.3 and 3C~303.1 show the joined effect of the two processes. The sources have undergone violent shock processes as a result of jet-cloud interaction, whose outflows were found in the innermost regions \citep[][and Fig.~\ref{VF}]{holt08}. In the case of 3C~303.1 we also detect turbulence in the large-scale gas. The jet-cloud interaction have triggered shock waves that, according to observations, not only alter the physical conditions of the environment but also create a local ionizing field. Nonetheless, the spectroscopical signatures found in both sources point toward the recombination region behind the shock as the main contribution to the emergent spectra, rather than the precursor. The gas compression as well as the presence of a non-negligible magnetic field create suitable conditions to improve the photoionizing power of the AGN radiation field. Therefore, we have found the way in which (kinematically acceptable) shock-ionization combines with AGN photoionization to trace the intriguing spectral characteristics observed in the EELR of these two CSS sources.

A brief description of the individual sources, along with the successes and failures of each model, is summarized in the following paragraphs.

3C~268.3 is not well fitted by shock-ionization: predictions underestimate the data when neon lines are combined, or a fake fitting is achieved in the region of most violent shocks which has no kinematical support. However, shocks must be present in order to explain the spatial behaviour of some emission lines.  We obtain a very good fitting with photoionization models when [NeV]$\lambda3426$ is involved, but even in these cases photoionization cannot account for the observed intensities by itself. Photoionization cannot account for the whole observed spectral characteristics since the relations amongst [OII]$\lambda3727$, [NeV]$\lambda3426$, U, and the distance from the source ($r$) completely contradict its hypothesis. Finally, given the ionization potential of [NeV] and HeII, the distances up to which we find the emission of these lines cannot be explained by the ionization structure of photoionized plasmas. So we conclude that shock processes must be highly important in order to explain the relations among [OII], [NeV], HeII, U, and $r$. Since \citet{holt08} have reported the existence of circumnuclear outflows in 3C~268.3, and this is the only region where our velocity field shows some (low-level) perturbation, the ionizing shocks are certainly triggered by that jet-cloud interaction. However, the contribution from photoionization is far from being negligible.

3C~303.1, in turn, is almost well explained by both scenarios, but the velocity field shows clear signatures of perturbation in the EELR. This result agrees with those of \citet{holt08}. Photoionization can reproduce the observations, but it fails in reproducing [NeV]$\lambda3426$. Shock-ionization predicts the observed line-intensities (line-ratios) for velocities in the range 300-500~km~s$^{-1}$. Consistently, the velocity field shows turbulent motions within the same velocity range. But, despite this important finding, the models also fail in reproducing [NeV]$\lambda3426$, so none of the two sets of models can account for this emission on their own. From our kinematical results, and those of \citet{holt08}, we confirm that shocks are present. And our detailed spectroscopical analysis has demonstrated that the EELR is shock-ionized. Nonetheless, from the analysis of the effect that an external fields has in shock-ionization predictions, we conclude that AGN photoionization is also required to explain, in combination with shock-ionization, the strength, the behaviour, and the spatial extent of the highest excitation line in the spectra: [NeV]$\lambda3426$. This combination of processes might have place in an scenario such as that proposed by \citet{shih13}, although our data do not show different patterns between the NW and SE regions.

\bsp
%
%

\section*{Acknowledgments}
VR thank the Support Staff of the Gemini Observatory for their help in the reduction process. We also want to thank to the anonymous referee for helping us to improve the presentation of the paper.
%
%

\bibliographystyle{mn2e}
\bibliography{biblios2.bib}

\begin{thebibliography}{}

\bibitem[\protect\citeauthoryear{{Allen}, {Groves}, {Dopita}, {Sutherland} \&
  {Kewley}}{{Allen} et~al.}{2008}]{all08}
{Allen} M.~G.,  {Groves} B.~A.,  {Dopita} M.~A.,  {Sutherland} R.~S.,
  {Kewley} L.~J.,  2008, \apjs, 178, 20

\bibitem[\protect\citeauthoryear{{Axon}, {Capetti}, {Fanti}, {Morganti},
  {Robinson} \& {Spencer}}{{Axon} et~al.}{2000}]{axon00}
{Axon} D.~J.,  {Capetti} A.,  {Fanti} R.,  {Morganti} R.,  {Robinson} A.,
  {Spencer} R.,  2000, \aj, 120, 2284

\bibitem[\protect\citeauthoryear{{Baldwin}, {Phillips} \&
  {Terlevich}}{{Baldwin} et~al.}{1981}]{bal81}
{Baldwin} J.~A.,  {Phillips} M.~M.,    {Terlevich} R.,  1981, \pasp, 93, 5

\bibitem[\protect\citeauthoryear{{Binette}, {Wilson}, {Raga} \&
  {Storchi-Bergmann}}{{Binette} et~al.}{1997}]{bin97}
{Binette} L.,  {Wilson} A.~S.,  {Raga} A.,    {Storchi-Bergmann} T.,  1997,
  \aap, 327, 909

\bibitem[\protect\citeauthoryear{{Binette}, {Wilson} \&
  {Storchi-Bergmann}}{{Binette} et~al.}{1996}]{bin96}
{Binette} L.,  {Wilson} A.~S.,    {Storchi-Bergmann} T.,  1996, \aap, 312, 365

\bibitem[\protect\citeauthoryear{{Cardelli}, {Clayton} \& {Mathis}}{{Cardelli}
  et~al.}{1989}]{cardelli89}
{Cardelli} J.~A.,  {Clayton} G.~C.,    {Mathis} J.~S.,  1989, \apj, 345, 245

\bibitem[\protect\citeauthoryear{{de Koff}, {Baum}, {Sparks}, {Biretta},
  {Golombek}, {Macchetto}, {McCarthy} \& {Miley}}{{de Koff}
  et~al.}{1996}]{koff96}
{de Koff} S.,  {Baum} S.~A.,  {Sparks} W.~B.,  {Biretta} J.,  {Golombek} D.,
  {Macchetto} F.,  {McCarthy} P.,    {Miley} G.~K.,  1996, \apjs, 107, 621

\bibitem[\protect\citeauthoryear{{de Vries}, {O'Dea}, {Baum} \& {Barthel}}{{de
  Vries} et~al.}{1999}]{deVries99}
{de Vries} W.~H.,  {O'Dea} C.~P.,  {Baum} S.~A.,    {Barthel} P.~D.,  1999,
  \apj, 526, 27

\bibitem[\protect\citeauthoryear{{de Vries}, {O'Dea}, {Baum}, {Sparks},
  {Biretta}, {de Koff}, {Golombek}, {Lehnert}, {Macchetto}, {McCarthy} \&
  {Miley}}{{de Vries} et~al.}{1997}]{deVries97}
{de Vries} W.~H.,  {O'Dea} C.~P.,  {Baum} S.~A.,  {Sparks} W.~B.,  {Biretta}
  J.,  {de Koff} S.,  {Golombek} D.,  {Lehnert} M.~D.,  {Macchetto} F.,
  {McCarthy} P.,    {Miley} G.~K.,  1997, \apjs, 110, 191

\bibitem[\protect\citeauthoryear{{Dopita} \& {Sutherland}}{{Dopita} \&
  {Sutherland}}{1995}]{dop95}
{Dopita} M.~A.,  {Sutherland} R.~S.,  1995, \apj, 455, 468

\bibitem[\protect\citeauthoryear{{Dopita} \& {Sutherland}}{{Dopita} \&
  {Sutherland}}{1996}]{dop96}
{Dopita} M.~A.,  {Sutherland} R.~S.,  1996, \apjs, 102, 161

\bibitem[\protect\citeauthoryear{{Fanaroff} \& {Riley}}{{Fanaroff} \&
  {Riley}}{1974}]{fr74}
{Fanaroff} B.~L.,  {Riley} J.~M.,  1974, \mnras, 167, 31P

\bibitem[\protect\citeauthoryear{{Feinstein}, {Macchetto}, {Martel}, {Sparks}
  \& {McCarthy}}{{Feinstein} et~al.}{1999}]{fei99}
{Feinstein} C.,  {Macchetto} F.~D.,  {Martel} A.~R.,  {Sparks} W.~B.,
  {McCarthy} P.~J.,  1999, \apj, 526, 623

\bibitem[\protect\citeauthoryear{{Fu} \& {Stockton}}{{Fu} \&
  {Stockton}}{2006}]{fu06}
{Fu} H.,  {Stockton} A.,  2006, \apj, 650, 80

\bibitem[\protect\citeauthoryear{{Fu} \& {Stockton}}{{Fu} \&
  {Stockton}}{2008}]{fu08}
{Fu} H.,  {Stockton} A.,  2008, \apj, 677, 79

\bibitem[\protect\citeauthoryear{{Groves} \& {Allen}}{{Groves} \&
  {Allen}}{2010}]{gro10}
{Groves} B.~A.,  {Allen} M.~G.,  2010, \na, 15, 614

\bibitem[\protect\citeauthoryear{{Holt}, {Tadhunter} \& {Morganti}}{{Holt}
  et~al.}{2008}]{holt08}
{Holt} J.,  {Tadhunter} C.,    {Morganti} R.,  2008, \mnras, 387, 639

\bibitem[\protect\citeauthoryear{{Holt}, {Tadhunter} \& {Morganti}}{{Holt}
  et~al.}{2009}]{holt09}
{Holt} J.,  {Tadhunter} C.,    {Morganti} R.,  2009, \mnras, 400, 589

\bibitem[\protect\citeauthoryear{{Komossa} \& {Schulz}}{{Komossa} \&
  {Schulz}}{1997}]{kom97}
{Komossa} S.,  {Schulz} H.,  1997, \aap, 323, 31

\bibitem[\protect\citeauthoryear{{Labiano}, {O'Dea}, {Gelderman}, {de Vries},
  {Axon}, {Barthel}, {Baum}, {Capetti}, {Fanti}, {Koekemoer}, {Morganti} \&
  {Tadhunter}}{{Labiano} et~al.}{2005}]{labiano05}
{Labiano} A.,  {O'Dea} C.~P.,  {Gelderman} R.,  {de Vries} W.~H.,  {Axon}
  D.~J.,  {Barthel} P.~D.,  {Baum} S.~A.,  {Capetti} A.,  {Fanti} R.,
  {Koekemoer} A.~M.,  {Morganti} R.,    {Tadhunter} C.,  2005, \aap, 436, 493

\bibitem[\protect\citeauthoryear{{O'Dea}}{{O'Dea}}{1998}]{odea98}
{O'Dea} C.~P.,  1998, \pasp, 110, 493

\bibitem[\protect\citeauthoryear{{O'Dea}, {de Vries}, {Koekemoer}, {Baum},
  {Morganti}, {Fanti}, {Capetti}, {Tadhunter}, {Barthel}, {Axon} \&
  {Gelderman}}{{O'Dea} et~al.}{2002}]{odea02}
{O'Dea} C.~P.,  {de Vries} W.~H.,  {Koekemoer} A.~M.,  {Baum} S.~A.,
  {Morganti} R.,  {Fanti} R.,  {Capetti} A.,  {Tadhunter} C.,  {Barthel} P.~D.,
   {Axon} D.~J.,    {Gelderman} R.,  2002, \aj, 123, 2333

\bibitem[\protect\citeauthoryear{{Penston}, {Robinson}, {Alloin},
  {Appenzeller}, {Aretxaga}, {Axon}, {Baribaud}, {Barthel}, {Baum}, {Boisson}
  \& et al.}{{Penston} et~al.}{1990}]{pen90}
{Penston} M.~V.,  {Robinson} A.,  {Alloin} D.,  {Appenzeller} I.,  {Aretxaga}
  I.,  {Axon} D.~J.,  {Baribaud} T.,  {Barthel} P.,  {Baum} S.~A.,  {Boisson}
  C.,    et al. 1990, \aap, 236, 53

\bibitem[\protect\citeauthoryear{{Peterson}}{{Peterson}}{1997}]{pet97}
{Peterson} B.~M.,  1997, {An Introduction to Active Galactic Nuclei}.
Cambridge, New York Cambridge University Press

\bibitem[\protect\citeauthoryear{{Privon}, {O'Dea}, {Baum}, {Axon}, {Kharb},
  {Buchanan}, {Sparks} \& {Chiaberge}}{{Privon} et~al.}{2008}]{pri08}
{Privon} G.~C.,  {O'Dea} C.~P.,  {Baum} S.~A.,  {Axon} D.~J.,  {Kharb} P.,
  {Buchanan} C.~L.,  {Sparks} W.,    {Chiaberge} M.,  2008, \apjs, 175, 423

\bibitem[\protect\citeauthoryear{{Reynaldi} \& {Feinstein}}{{Reynaldi} \&
  {Feinstein}}{2013a}]{reynaldi13a}
{Reynaldi} V.,  {Feinstein} C.,  2013a, \mnras, 430, 2221

\bibitem[\protect\citeauthoryear{{Reynaldi} \& {Feinstein}}{{Reynaldi} \&
  {Feinstein}}{2013b}]{reynaldi13b}
{Reynaldi} V.,  {Feinstein} C.,  2013b, \mnras, 435, 1350

\bibitem[\protect\citeauthoryear{{Robinson}, {Tadhunter} \& {Dyson}}{{Robinson}
  et~al.}{2002}]{rob02}
{Robinson} T.~G.,  {Tadhunter} C.,    {Dyson} J.~E.,  2002, \mnras, 331, L13

\bibitem[\protect\citeauthoryear{{Rosario}, {Whittle}, {Nelson} \&
  {Wilson}}{{Rosario} et~al.}{2010}]{ros10a}
{Rosario} D.~J.,  {Whittle} M.,  {Nelson} C.~H.,    {Wilson} A.~S.,  2010,
  \apj, 711, L94

\bibitem[\protect\citeauthoryear{{Shapiro}, {Clocchiatti} \& {Kang}}{{Shapiro}
  et~al.}{1992}]{shapiro92}
{Shapiro} P.~R.,  {Clocchiatti} A.,    {Kang} H.,  1992, \apj, 389, 269

\bibitem[\protect\citeauthoryear{{Shih}, {Stockton} \& {Kewley}}{{Shih}
  et~al.}{2013}]{shih13}
{Shih} H.-Y.,  {Stockton} A.,    {Kewley} L.,  2013, \apj, 772, 138

\bibitem[\protect\citeauthoryear{{Sol{\'o}rzano-I{\~n}arrea}, {Tadhunter} \&
  {Axon}}{{Sol{\'o}rzano-I{\~n}arrea} et~al.}{2001}]{sol01}
{Sol{\'o}rzano-I{\~n}arrea} C.,  {Tadhunter} C.,    {Axon} D.~J.,  2001,
  \mnras, 323, 965

\bibitem[\protect\citeauthoryear{{Stasi{\'n}ska}}{{Stasi{\'n}ska}}{1984}]{stas84}
{Stasi{\'n}ska} G.,  1984, \aap, 135, 341

\bibitem[\protect\citeauthoryear{{Stasi{\'n}ska}}{{Stasi{\'n}ska}}{2009}]{stas09}
{Stasi{\'n}ska} G.,  2009, in {Cepa} J.,  ed., Canary Islands Winter School on
  Astrophysics, 18th, 2006, The Emission-Line Universe. Instituto de
  Astrof{\'{\i}}sica de Canarias: Canary Islands Winter School on Astrophysics,
  {What can emission lines tell us?}

\bibitem[\protect\citeauthoryear{{Tadhunter}, {Fosbury} \& {Quinn}}{{Tadhunter}
  et~al.}{1989}]{tad89}
{Tadhunter} C.,  {Fosbury} R.~A.~E.,    {Quinn} P.~J.,  1989, \mnras, 240, 225

\end{thebibliography}
\newpage




\begin{table}
\begin{center}
\caption{Set of emission-lines in the EELR of 3C~268.3 and 3C~303.1. Fluxes are relative to that of H$\beta$ for each selected position.}
\label{table-1}
\begin{tabular}{lcccccc}
\hline
\hline
\multicolumn{7}{c}{3C 268.3}\\
\hline
\hline
 & \multicolumn{6}{c}{Flux}\\
Line (\AA) & \multicolumn{3}{c}{Northwest (NW)} & \multicolumn{3}{c}{Southeast (SE)} \\
\hline
 & -1.6'' & -0.7'' & -0.3'' & 0.6''& 1.0'' & 2.1'' \\
\hline

[Ne {\sc v}]$\lambda3426$ & - & - & - & 0.32 & 0.75 & 1.55 \\

[O {\sc ii}]$\lambda3727$ & 0.8 & 1.16 & 0.94 & 2.05 & 2.7 & 1.96 \\

[Ne {\sc iii}]$\lambda3869$ & - & 0.28 & 0.22 & 0.68 & 0.75 & 1.07 \\

[Ne {\sc iii}]$\lambda3968$ & - & 0.29 & 0.08 & 0.21 & 0.17 & 0.49 \\

H$\delta$ (4101) & - & 0.18 & 0.06 & 0.03 & 0.09 & 0.32 \\

H$\gamma$ (4340) & - & 0.18 & 0.14 & 0.22 & 0.23 & 0.41 \\

[O {\sc iii}]$\lambda4363$ & - & 0.09 & - & 0.15 & - & - \\

He {\sc ii} (4686) & - & 0.14 & 0.17 & 0.14 & 0.25 & 0.32 \\

H$\beta$ (4861) & 1 & 1 & 1 & 1 & 1 & 1 \\

[O {\sc iii}]$\lambda4959$ & 2.28 & 5.06 & 3.53 & 3.24 & 3.69 & 2.88 \\

[O {\sc iii}]$\lambda5007$ & 4.0 & 11.93 & 8.71 & 9.55 & 11.61 & 8.46 \\

\hline
\hline
\multicolumn{7}{c}{3C 303.1}\\
\hline
\hline
 & \multicolumn{6}{c}{Flux}\\
Line (\AA) & \multicolumn{3}{c}{Northwest (NW)} & \multicolumn{3}{c}{Southeast (SE)} \\
\hline
 & -1.6'' & -1.02'' & -0.6'' & 0.6''& 1.2'' & 1.7'' \\
\hline

[Ne {\sc v}]$\lambda3426$ & - & - & - & 0.12 & 0.4 & - \\

[O {\sc ii}]$\lambda3727$ & 6.19 & 5.37 & 4.01 & 3.6 & 3.11 & 4.31 \\

[Ne {\sc iii}]$\lambda3869$ & 1.36 & 1.01 & 0.77 & 0.58 & 0.61 & 0.83 \\

H {\sc i}+He {\sc i} (3890) & - & - & 0.23 & 0.15 & 0.22 & - \\

[Ne {\sc iii}]$\lambda3968$ & - & 0.48 & 0.39 & 0.28 & 0.19 & - \\

H$\delta$ (4101) & - & - & 0.19 & 0.21 & - & - \\

H$\gamma$ (4340) & - & 0.61 & 0.52 & 0.37 & 0.33 & 0.37 \\

[O {\sc iii}]$\lambda4363$ & - & - & 0.18 & 0.14 & 0.13 & - \\

He {\sc ii} (4686) & - & 0.31 & 0.19 & 0.12 & 0.13 & - \\

H$\beta$ (4861) & 1 & 1 & 1 & 1 & 1 & 1 \\

[O {\sc iii}]$\lambda4959$ & 3.29 & 3.61 & 3.01 & 2.06 & 2.39 & 3.62 \\

[O {\sc iii}]$\lambda5007$ & 10.14 & 11.27 & 9.68 & 6.11 & 7.11 & 10.52 \\

\hline
\end{tabular}
\end{center}
\medskip

The emission-line labeled as [Ne {\sc iii}]$\lambda3968$ is actually a complex formed by the following emission-lines: [Ne {\sc iii}]$\lambda3967.4$, Ca {\sc ii}(3968.44), and He~{\sc ii}(3968.43). Since the order reflects the expected intensity, we choose to label the complex as the putative more intense emission-line.

\end{table}

\begin{table}
\begin{center}
\caption{Differences among mix-medium photoionization models.}
\label{table-2}
\begin{tabular}{cccc}
\hline
\hline
Model & & U & $n_e$ [cm$^{-3}$] \\
\hline
T & MB & 0.04 & 50 \\
  & IB & $5.2 \times 10^{-4}$ & $2.3\times 10^3$ \\
H & MB & 0.5 & $1 \times 10^3$ \\
  & IB & $6.5 \times 10^{-4}$ & $\gtrsim 1.2 \times 10^4$ \\
M & MB & 0.05 & $1 \times 10^3$ \\
  & IB & $6.5 \times 10^{-4}$ & $\gtrsim 1.2 \times 10^4$ \\
L & MB & 0.02 & $1 \times 10^3$ \\
  & IB & $6.5 \times 10^{-4}$ & $\gtrsim 1.2 \times 10^4$ \\
\hline
\end{tabular}
\end{center}
\end{table}

\newpage

\newpage

\newpage


\label{lastpage}

\end{document}